\documentclass[journal=jacsat,manuscript=article]{achemso}

\usepackage[version=3]{mhchem} 
\usepackage{braket}
\usepackage{xcolor}
\usepackage[normalem]{ulem}

\author{Juan E. Arias-Martinez}
\email{juanes@berkeley.edu}
\affiliation
{{Kenneth S. Pitzer Center for Theoretical Chemistry, Department of Chemistry, University of California, Berkeley, California 94720, USA}}
\alsoaffiliation{Chemical Sciences Division, Lawrence Berkeley National Laboratory, Berkeley, California 94720, USA}
\author{Leonardo A. Cunha}
\affiliation
{{Kenneth S. Pitzer Center for Theoretical Chemistry, Department of Chemistry, University of California, Berkeley, California 94720, USA}}
\alsoaffiliation{Chemical Sciences Division, Lawrence Berkeley National Laboratory, Berkeley, California 94720, USA}
\author{Katherine J. Oosterbaan}
\affiliation{{Kenneth S. Pitzer Center for Theoretical Chemistry, Department of Chemistry, University of California, Berkeley, California 94720, USA}}
\author{Joonho Lee}
\affiliation{Department of Chemistry, Columbia University, New York, New York 10027, USA}
\author{Martin Head-Gordon}
\email{mhg@cchem.berkeley.edu}
\affiliation
{{Kenneth S. Pitzer Center for Theoretical Chemistry, Department of Chemistry, University of California, Berkeley, California 94720, USA}}
\alsoaffiliation{Chemical Sciences Division, Lawrence Berkeley National Laboratory, Berkeley, California 94720, USA}

\title[An \textsf{achemso} demo]
  {Accurate core excitation and ionization energies from a state-specific 
  coupled-cluster singles and doubles approach}

\abbreviations{HF, SCF, CCSD, BSL}

\begin{document}

\begin{tocentry}

\centering
\includegraphics[width=0.80\linewidth]{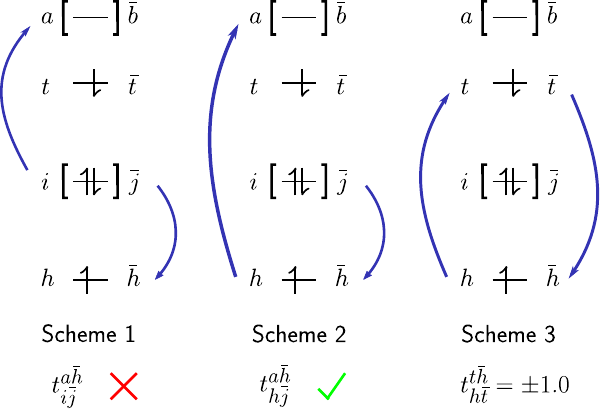}.

\end{tocentry}

\begin{abstract}
We investigate the use of orbital-optimized references in 
conjunction with single-reference coupled-cluster theory with
single and double substitutions (CCSD) for the study of core
excitations and ionizations of 18 small organic molecules, 
without the use of response theory or equation-of-motion (EOM) formalisms. 
Three schemes are employed to successfully address the convergence 
difficulties associated with the coupled-cluster equations, and the 
spin contamination resulting from the use of a spin symmetry-broken 
reference, in the case of excitations. In order to gauge the inherent 
potential of the methods studied, an effort is made to 
provide reasonable basis set limit estimates for the 
transition energies. Overall, we find that the two best-performing schemes 
studied here for $\Delta$CCSD are capable of predicting excitation and 
ionization energies with errors comparable to experimental accuracies.
The proposed $\Delta$CCSD schemes reduces statistical 
errors against experimental excitation energies by more than a factor 
of two when compared to the frozen-core core-valence separated (FC-CVS) 
EOM-CCSD approach - a successful variant of EOM-CCSD tailored towards 
core excitations.
\end{abstract}

\section{Introduction}
Recent decades have seen improvements in the handling of high-energy 
ultra-violet (XUV) and X-ray light in terms of coherence,
\cite{Yu2019, Halavanau2020}
intensity, 
\cite{Halavanau2020}
and time control. 
\cite{KumarMaroju2005, Duris2019, Halavanau2020}
As a result, scientists have been able to observe 
phenomena in chemistry,
\cite{Chang2020, Loh2020, Lin2021}
material sciences,
\cite{Sette1998, Attar2020}
and physics
\cite{Mazza2020, Haynes2021}
that were previously inaccessible.
Furthermore, the increasing availability of table-top equipment
\cite{Popmintchev2012, Zimmermann2020, Barreau2020, Scutelnic2021} 
capable of generating the light required for core spectroscopies has 
extended the use of said techniques for a variety of new studies.\cite{Epshtein2020} 
Efficiently and accurately modeling core excited states presents 
challenges that a useful methodology should address, chief among them 
the large charge rearrangement associated with the creation of the core hole. 
Within the independent particle model, this charge rearrangement results 
in a strong contraction of the orbitals due to the decreased nuclear 
screening - this is referred to as orbital relaxation in the literature.
The most widely used method for calculating valence excited states, 
time-dependent density functional theory (TD-DFT), struggles to 
describe core excited states (and charge transfer states in general) 
because the linear-response (LR) formalism fails to account for 
the charge rearrangement when standard exchange-correlation 
functionals are used.\cite{BEsley2010, Lyon2017, Maitra2017, Maitra2021}
This failure of TD-DFT has been associated 
with the self-interaction error and/or the difficulty 
of describing orbital relaxation, and it manifests through 
errors on the order of tens of eV for the K-edge of main 
group with larger errors being observed for heavier elements.
\cite{Dreuw2004, Hait2020-core_excitations, Cunha2022}
When used to assess experimental results, the TD-DFT spectra is often shifted 
to either match the first or the brightest feature, or else to have it match 
the experimental results as much as possible.\cite{Scutelnic2021} 
Progress in the use of LR-based DFT for core spectroscopies has been 
made through the 
design of functionals specialized to core excitations.\cite{BEsley2010}\\

To circumvent the uncertainty associated with the choice of functionals, 
established wave function theories that are well-regarded for their accuracy in 
describing valence excitations, such as EOM-CC
theory and algebraic diagrammatic construction (ADC), have 
been extended to core excitations by implementing techniques that target
the high-energy roots of their effective Hamiltonians. 
\cite{Cederbaum1980, Barth1981, Barth1985, Coriani2012, Kauczor2013, Coriani2015, Wenzel2015, Vidal2019}
A challenge that some techniques implemented in the last decade 
faced\cite{Coriani2012, Kauczor2013}
is the fact that core excited states are resonances 
embedded in an Auger continuum. The earlier idea of the core-valence 
separation (CVS),\cite{Cederbaum1980, Barth1981} where the 
continuum is explicitly decoupled from the core excited states 
in some way, emerged as a successful solution to the
problem, and therefore as the preferred protocol to 
target core excitations.\cite{Barth1985, Coriani2015, Vidal2019}
It is worth noting, however, that the details 
of the CVS implementation may lead to differences on the order of 
eVs.\cite{Vidal2019}

An alternative approach followed by state-specific methods such as $\Delta$SCF
\cite{Triguero1999, Hait2020}
and its correlated relatives, 
\cite{Besley2009, Duflot2010, Shim2011, Ljubic2014, Zheng2019, Zheng2020, Lee2019, Huang2021}
the closely-related Transition Potential (TP)-SCF approaches,\cite{Hu1996, Triguero1998, Triguero1999, Michelitsch2019}
a number of multi-reference (MR) wave function models,
\cite{Brabec2012, Sen2013, Dutta2014, Maganas2019}
excited state mean field theory,\cite{Garner2020-2} 
and Monte-Carlo-based approaches,\cite{Garner2020-1} 
is to account for relaxation in some way by optimizing for a target 
state.  The $\Delta$SCF approach, for example, converges a set of orbitals
in a configuration that resembles the one-electron picture of the 
core excitation in question. These are non-Aufbau solutions to the 
self-consistent field (SCF) equations and are often saddle points in 
orbital space. Similarly, TP-SCF employs configurations optimized 
for a fractional core occupancy in the hopes of providing 
a reference of similar quality for both the ground and the core 
excited states. A difficulty of orbital-optimized excited state 
approaches is the possibility of landing on an undesired SCF 
solution of lower energy. In the context of mean-field approaches, 
such as Hartree-Fock (HF) and density functional theory (DFT), 
this issue has been addressed by algorithms specialized for 
excited state optimization, such as the maximum overlap method 
(MOM)\cite{Gilbert2008}, and, more recently, the initial MOM (IMOM)\cite{barca2018simple}, square-gradient 
minimization (SGM)\cite{Hait2020} and state-targeted energy 
projection (STEP)\cite{Carter-Fenk2020} methods. 

$\Delta$SCF has been used for decades to calculate core ionizations 
with success.\cite{Triguero1999, Besley2009, Ljubic2014} In the cases 
where there are symmetry-equivalent atoms present in the 
system, an orbital localization procedure (such as that of Boys\cite{Boys1960}) 
must be carried out on the core orbitals prior to SCF re-optimization 
to allow for proper orbital relaxation.\cite{Nooijen1995, Zheng2019, Brumboiu2022} 
The spatial symmetry breaking technically renders these situations 
multi-reference (MR) since multiple configurations must be 
re-combined via non-orthogonal configuration interaction (NOCI) 
to yield states of the proper spatial symmetry. In practice, the 
splitting between the symmetry-adapted configurations is small,
\cite{Liu2019, Oosterbaan2020} so that the MR character associated 
with the core hole localization can be disregarded without serious 
error. The $\Delta$SCF ionization energies, as calculated with the 
spatially symmetry-broken configurations are often good estimates 
of what would be observed in an experiment.

Studies on core excitations with $\Delta$SCF have been more sparse until
recently.\cite{Besley2009} In some measure this is due to 
the fact that MR character now factors in because of the need for two 
configurations for a spin-pure description of the excited state. The 
approximate spin-projection scheme (AP) established a way to estimate 
the excitation energy of the pure singlet, provided that the energies of a 
spin-contaminated singlet and the pure triplet are known.\cite{yamaguchi1988spin,Kitagawa2009} 
An attractive alternative to AP for $\Delta$SCF calculations is the use of restricted 
open-shell Kohn-Sham orbitals (ROKS), which optimizes the spin-pure singlet 
energy as computed via the AP scheme for a mixed and a triplet 
(M$_s$ = 1) configuration sharing the same set of restricted open-shell 
(RO) orbitals.\cite{frank1998molecular,Filatov1999,kowalczyk2013excitation,hait2021orbital} 
Recently, this technique (and a generalized version for radicals
\cite{hait2020accurate}) has been used to study core excited states 
with the best-performing functional (SCAN) achieving an impressive 
0.2 eV root-mean-squared-deviation (RMSD) from experimental results 
for a representative set of small organic molecules.
\cite{Hait2020-core_excitations} 
With an appropriate treatment of scalar relativistic 
effects, ROKS has also been employed to tackle the K-edge of third-group elements.\cite{Cunha2022}

Excited SCF solutions are often a better reference than the ground state 
for finding alternative solutions to the CC equations, which in turn
are reasonable approximations to the true excited states.\cite{Meissner1993} 
Explicit SCF re-optimization takes care of the strong orbital 
relaxation and allows single-reference (SR) post-HF methods 
such as second order M{\o}ller-Plesset perturbation theory (MP2) 
and CC to focus on addressing 
the remaining dynamic correlation of a system. Core ionized 
states of closed-shell systems are perfect cases to be treated 
by these models and they have been studied via $\Delta$MP2\cite{Triguero1999, Besley2009, Duflot2010, Shim2011, Ljubic2014}
and, more recently, $\Delta$CCSD(T) \cite{Zheng2019, Lee2019, Zheng2020}. 
The last decade has seen an effort to also employed 
explicitly-relaxed orbitals on a (wave-function-based) correlated 
calculation for singlet excited states
\cite{Besley2009, Lee2019, Matthews2020, Huang2021, Brabec2012, Sen2013, Dutta2014, Maganas2019}. 
Among these, the wave function theories employing explicit MR construction 
often constrain them to study few molecules in small basis sets, 
which means they can only be compared to other computational methods in the same small basis sets.
\cite{Brabec2012, Sen2013, Dutta2014, Maganas2019}
Simons and Matthews have recently proposed a theory, TP CC, that 
employs a TP SCF reference for an EOM-CC calculation of 
the core excited states.\cite{Simons2021} This model inherits some of the 
advantages of both state-specific methods - orbital relaxation - 
while retaining the advantages of EOM-CC: inherent spin-adaptation 
of the excited states, a full spectrum with a single calculation, 
and straightforward transition properties. The cost to pay comes from 
relying on a deteriorated description 
of the ground state relative to standard CC, controlled by tuning the fractional 
occupation number of the core orbital. Even though this renders the model 
arbitrary, to some extent, Simons and Matthews have carried out a study 
to find an optimal core occupancy parameter transferable across edges 
of the same element, making this a promising method for reliable and 
affordable high-accuracy wave function X-ray calculations.\cite{simons2022transitionpotential}

Owing to the simple nature of the MR character of singly core excited 
states of closed shell systems (namely, a two-determinant CSF) the 
objective of this paper is to assess the use of SR CC formalism 
(limited to the level of singles and doubles - CCSD) with orbital-optimized 
references for the prediction of core excited state energies.
We believe the schemes proposed and analyzed in the present work
could be useful for providing theoretical benchmark numbers for 
core excited and ionized states. As observed in this study, 
the best $\Delta$CCSD models significantly outperforms
FC-CVS-EOM-CCSD while retaining its $\mathcal{O}(N^6)$ scaling, 
with N being the size of the one-electron basis set employed. 
Furthermore, unlike FC-CVS-EOM-CC, it does not rely on 
cancellation of errors.\cite{Vidal2019} Per previous studies, 
the formally-correct CVS-EOM-CC implementation\cite{Coriani2012} 
is likely to require full triples, i.e. $\mathcal{O}(N^8)$-
scaling CVS-EOM-CCSDT, to reach similar accuracy.\cite{Liu2019, Matthews2020}
In contrast, the protocols presented here are well-defined in that 
only the molecule and the transition of interest needs to be 
specified - the proper ground state CC wave function and energies 
are used as is, unlike in the FC-CVS approach or the TP CC
method, and no compromise in the excited state wave function is 
made either. 

In early work along the lines of $\Delta$CCSD, where Nooijen and 
Bartlett employed a relaxed core-ionized reference for a 
subsequent electron-attachment (EA) EOM-CCSD for the calculation 
of core excited states, they recognized two major challenges related 
to these sort of calculations.\cite{Nooijen1995} The first is 
how to treat the electron correlation effects that couple core orbitals 
with either other core levels or valence levels. De-excitation into the 
core hole can lead either to numerical instabilities or variational 
collapse towards the ground state. Therefore a suitable adaptation 
of SR CCSD for state-specific optimization of core excited states must treat core 
correlation, as well as removing potentially ill-behaved amplitudes. 
The second is the issue of ensuring proper spin symmetry in the final 
CC wave function.

This paper is organized as follows. After a review of the appropriate theory, 
we describe three candidate approaches that we deem potentially promising. 
Two of them employ Yamaguchi's AP approach\cite{yamaguchi1988spin}, while
the third one instead enforces correct spin symmetry at the ROHF level by 
constraining the amplitude of the double substitution that flips the spins 
of the two half-occupied orbitals to $+1$ for singlet and $-1$ for triplet 
states. A comparison of these approaches is then made against 
successful core excited state theories, ROKS(SCF) and FC-CVS-EOM-CC, with 
the ultimate judge being the experimental results. The energetic differences 
between the singlet and triplet core excited states, presumed to be accurate 
enough to make a statement about them, are presented. An effort is made to 
reach basis set convergence for all methods in order to exclude this 
factor from the discussion as much as possible and focus
on their inherent performance. Despite the computational demands 
of approaching the basis set limit (BSL) for CC methods constraining us 
to molecules with at most two heavy atoms, the data set is 
diverse in terms of the elements targeted (Be, C, N, O, F, Ne) 
and in terms of the excited state character ($\sigma^*$, $\pi^*$, 
Rydberg). In total, a set of 21 excitations and 18 
ionizations on 18 small closed-shell organic molecules is used.

    We emphasize that our focus is on reporting excitation energies obtained through different proposed schemes within the $\Delta$CC framework. At present, our work does not extend the discussion of $\Delta$CC to compute transition properties. Obtaining such properties would be cumbersome and expensive due to, in part, the use of different sets of amplitudes for the bra and ket CC states. As pointed out in Ref. \citenum{Lee2019}, a potentially useful strategy to circumvent this exponential cost would be to use linearized wave functions obtained from the CC amplitudes from either the ground or core excited states, but we did not explore this further in our study.

\subsection{Background}
Following convention, we will reserve the indexes $i, j, k \dots$ for any occupied orbital, 
$a, b, c \dots$ for any virtual orbital, and $p, q, r \dots$ for an arbitrary 
orbital.
For the CCSD amplitudes, we will use the symbols 
$t_i^a$ and $t_{ij}^{ab}$, collected in $T_1$ and 
$T_2$. 

For a set of orbitals that are not necessarily canonical, the CCSD amplitude equations take the following form: 
\begin{align}
    D_{i}^{a} t_{i}^{a}     &= F_{ia} + w_{i}^{a}(T_1,T_2)  \label{eq:t1}\\
    D_{ij}^{ab} t_{ij}^{ab} &= \bra{ij}\ket{ab} + w_{ij}^{ab}(T_1,T_2) \label{eq:t2} 
\end{align}
The terms $w_{i}^{a}(T_1,T_2)$ and $w_{ij}^{ab}(T_1,T_2)$ in Eqs. \ref{eq:t1} and \ref{eq:t2} contain terms that are linear and higher in $T_1$ and $T_2$ separate from the orbital energy differences, $D_{i}^{a}$ and $D_{ij}^{ab}$ defined below\cite{Stanton1991}. 
\begin{align}
    D_{i}^{a} &= \varepsilon_i - \varepsilon_a \\
    D_{ij}^{ab} &= \varepsilon_i + \varepsilon_j - \varepsilon_a - \varepsilon_b
\end{align}
$\varepsilon_p$ are the orbital energies themselves. 
$D_{i}^{a}$ and $D_{ij}^{ab}$ will always be negative when employing a ground 
state reference and, in the absence of strong correlation, 
are large enough to make the $T$ amplitudes well behaved 
(i.e. $\mathrm{max}\left[|t_i^a|,|t_{ij}^{ab}| \right] \ll 1$). 
State-specific optimization of a core excited state, on the other 
hand, correlates a non-Aufbau SCF reference. Here, we make use of 
three different kinds of such (beta) core excited references: (i) 
open-shell, symmetry-broken $M_S=0$ references for the calculation 
of the singlet core excited states; open-shell, (ii) spin-pure triplet 
$M_S=1$ references for the AP approach, when needed; and (iii) 
open-shell, spin-pure $M_S=\frac{1}{2}$ doublet references for the 
calculation of core ionized states. In the case of the spin-pure 
triplet and pure doublet references, standard ROHF is used in 
conjunction with MOM. The use of unrestricted orbitals 
for the symmetry-broken reference was found to be detrimental to 
some of our $\Delta$CC schemes, so ROKS(HF) orbitals, 
followed by a Fock-build for the broken-symmetry singlet state and further 
pseudocanonicalization, were employed instead.

With these choices of reference, and specific to the case of 
core excitations, the presence of a virtual orbital 
with a large negative energy representing the core hole (we reserve the indexes $h$ and
$\bar{h}$ for the occupied alpha core orbital and the virtual beta core orbital) 
allows for denominators $D_i^a$ and $D_{ij}^{ab}$ to be positive 
when $a = \bar{h}$. In the case of single excitations, $a^\dag_{\bar{h}}a_i$, 
this occurs when the occupied orbital has a higher orbital energy than the 
core virtual
\begin{equation}\label{eq:single}
    \varepsilon_i > \varepsilon_{\bar{h}}
\end{equation}
The condition in Eq.~\ref{eq:single} holds unless there are other core orbitals of lower 
orbital energy. In the case of double excitations, 
$a^\dag_{\bar{h}}a_i a^\dag_b a_j$, $D_{ij}^{\bar{h}b}$ will be positive when
\begin{equation}\label{eq:double}
    \varepsilon_i + \varepsilon_j - \varepsilon_b > \varepsilon_{\bar{h}}
\end{equation}
One scenario where this happens is when the excitation 
$a^\dag_{\bar{h}} a_i$ involves a valence occupied orbital and 
the excitation $a^\dag_b a_j$  involves only valence orbitals.
\cite{Nooijen1995} The denominator $D_{ij}^{\bar{h}b}$ can still be 
negative if the other virtual has an orbital energy 
$\varepsilon_b$ positive and large enough to break Eq.~\ref{eq:double}. 
Furthermore, the orbital energies can conspire to make 
$\varepsilon_i + \varepsilon_j - \varepsilon_b \approx \varepsilon_{\bar{h}}$, 
rendering $D_{ij}^{ab} \approx 0$. 
Depending on the ability of the basis set to describe the 
high-lying virtual orbitals associated with the continuum, 
the denominator associated with double excitations can get 
arbitrarily close to zero, leading to numerical difficulties 
in solving for the T amplitudes (and of course divergence 
of perturbation theory methods, such as MP2). 

Close-to-zero denominators also yield numerical instabilities 
in the context of EOM-CC. In their study of EOM-CC-IP for K-edge 
ionization energies, Liu et al. found that spurious high-lying valence 
excited states that are quasi-degenerate with the core excited state 
result in erratically-converging correlation energies with respect to 
basis set.\cite{Liu2019}  The core-valence separation (CVS) scheme is a 
proposed solution to this numerical problem; in this approach, core 
excitations are excluded from the ground state amplitudes, and all-valence 
excitations are excluded from the EOM amplitudes.\cite{Vidal2019} 
The spurious couplings with the high-lying continuum 
excited states are then removed by design.

In a spirit similar to the CVS scheme, Zheng \emph{et al.}
proposed to exclude the virtual core orbital from the correlation 
treatment to address the divergence problem in the $\Delta$CC 
calculations of core ionizations.\cite{Zheng2019, Zheng2020} 
Some of us adopted a similar strategy where we freeze 
up the doubly-vacant core orbital all together when studying 
double-core excitations.\cite{Lee2019} Zheng \emph{et al.} 
found the missing correlation to be relevant for 
accurate core ionizations and uses estimates from fully-correlated 
CC calculations with decreasing denominator thresholds to account 
for it.

\section{Computational details}

A development version of Q-Chem 5.4\cite{Epifanovsky2021} 
was used for all calculations. 
Experimental geometries available on the NIST computational database
\cite{FIPS1402} were used throughout this work. 
An atomic relativistic correction calculated via the Douglas-Kroll-Hell 
method, found to be nearly independent of basis-set and molecule 
for the main group elements, is added to all calculations 
(0.012, 0.09, 0.18, 0.34, 0.57, and 0.91 eV for Be, C, N, O, F, and Ne.)\cite{Takahashi2017} 
For two of the three schemes of $\Delta$CC we employ, the 
calculated singlet excited states are spin contaminated; 
the AP method is used to estimate the spin-pure excitation 
energies. Aside from the amplitudes excluded in the 
different schemes, the CC calculations of both the ground and 
excited states are all-electron.

Our best attempt was made at comparing the excitation or ionization 
energies near their BSL values. To that end, different procedures 
involving specialized basis sets were employed for obtaining an 
approximate BSL for the different methods. The aug-pcX-3 (heavy)/ 
aug-pcseg-2 (hydrogen) basis was used to approximate the BSL for the 
ROKS(SCF) calculations.\cite{Ambroise2019} A (99, 590) Euler-Maclaurin-Lebedev 
grid was used for the computation of the exchange-correlation 
integrals for the ROKS(SCAN) calculations. The aug-ccX-nZ (heavy) / aug-cc-pVTZ (hydrogen) 
bases,\cite{Ambroise2021} extrapolated using the two-point X$^{-3}$ scheme\cite{Helgaker1997, halkier1998basis} 
with n = T, Q, were used to approximate the BSL for the EOM-CC calculations.
As noted in a recent study, such an extrapolation scheme is appropriate 
for core excitations via EOM-CC.\cite{Carbone2019}
All ROKS(SCF) and EOM-CC calculations were also run 
with the standard Dunning aug-cc-pCVXZ (X = D, T, Q) family of 
bases\cite{woon1995gaussian,peterson2002accurate} and a slower 
convergence towards a similar BSL value was observed (SI).

Of the basis sets available, none were designed with 
both explicit orbital relaxation via SCF and correlation with 
wave function methods in mind. We used the TQ-extrapolated
aug-cc-pCVXZ (heavy) / aug-cc-pVDZ (hydrogen) numbers as the best 
BSL estimate of the correlated $\Delta$ calculations.

The only exception to these choices of basis set was for the calculated 
Rydberg excitations in \ce{Ne}. As expected 
for a full-fledged Rydberg excitation, significant 
differences between the aug-cc-pCVXZ and its doubly-augmented 
counterparts were observed in this case. The BSL core excited 
states for this atom are given by the d-aug-cc-pCV5Z for ROKS(SCF), 
Q5-extrapolated d-aug-cc-pCVXZ for EOM-CC, and TQ-extrapolated 
d-aug-cc-pCVXZ for the correlated $\Delta$ methods. No severe difference of 
a similar sort was found in any other molecule studied in this data set, 
including the rest of the isoelectronic ten electron series (SI).

\section{Approaches to inclusion of core-valence correlation}

\subsection{Scheme 0: Using the full set of amplitudes}

To motivate the need for the schemes presented in the 
following subsections, we begin by exploring the behavior 
of the correlated methods with no modifications. The Fock matrix 
and MO coefficients of the optimized excited reference are passed to the 
correlated calculation and all amplitudes (e.g. all singles and doubles in CCSD) are included; we 
refer to this as Scheme 0 (S0). Scheme S0 would not be of use 
for real applications because of the possibility of variational collapse, and limitations
of today's standard iterative CC solvers. Nevertheless, it provides useful insight in the few 
cases where the coupled cluster equations do converge. Such systems 
are few-atom molecules in a small basis, where there are no orbitals 
of the right energy to make the denominators small enough. 

\begin{figure}
  \includegraphics{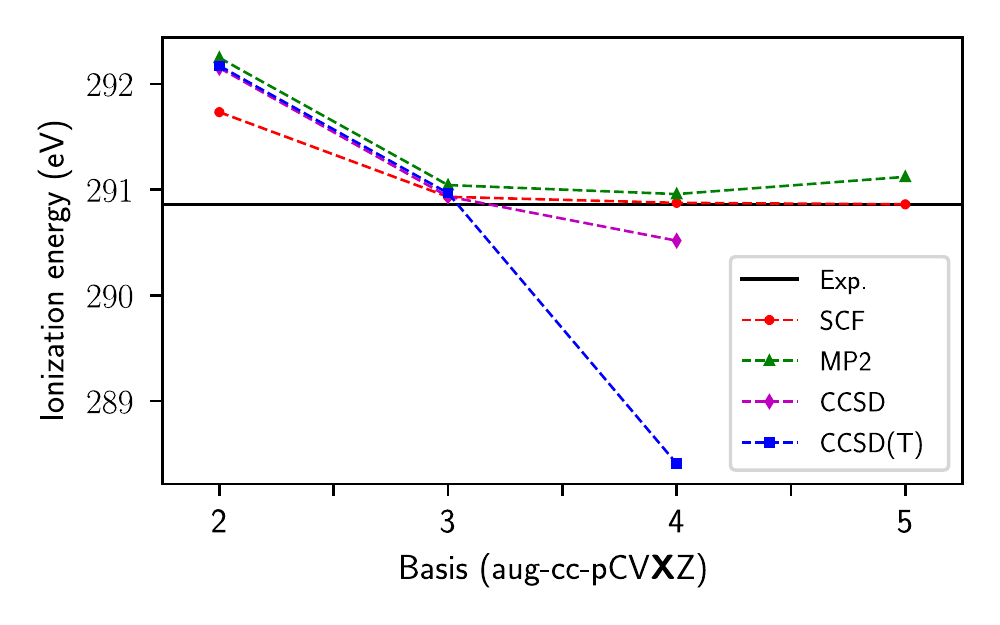}
  \caption{\ce{CH4} ionization at the Frank-Condon geometry}
  \label{fgr:CH4_ion}
\end{figure}

\begin{figure}
  \includegraphics{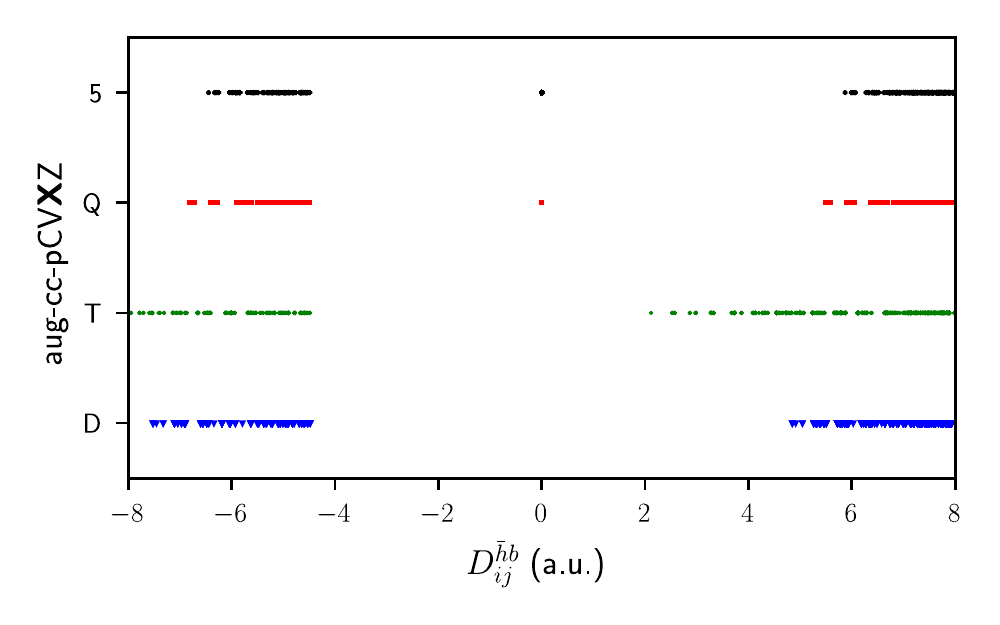}
  \caption{Some values for the denominators associated with excitation into the core virtual for the 
  \ce{CH4} core-ionized reference.}
  \label{fgr:denominators}
\end{figure}

Figure \ref{fgr:CH4_ion} shows the basis set convergence of the \ce{CH4} 
core ionization energies, as calculated with the $\Delta$-based methods, 
with respect to increasing cardinality of the aug-cc-pCVXZ basis set. 
The $\Delta$SCF values converge quickly, with the 5Z result decreasing 
the calculated ionization energy by only 0.014 eV from the 
QZ numbers. The results for all the correlated 
$\Delta$ methods are within 0.1 eV of each other up until the 
QZ level, where they begin to diverge. At the 5Z level, the CCSD 
equations fail to converge and the $\Delta$MP2 results break 
monotonicity. An analysis of the denominators associated with excitations into 
the core virtual (Figure \ref{fgr:denominators})
reveals that, for all basis sets, there are positive 
denominators and, furthermore, that a close-to-zero denominator 
appears at the QZ level. Once the complexity of the molecule 
increases, the virtual space will begin to populate the problematic 
orbital energy range associated with near-zero denominators even when using small basis sets. Yet the CCSD(S0) results, 
at the very least, suggest that accurate results via $\Delta$-based methods 
could be obtained if the irregularities caused by small denominators were addressed.

\subsection{Scheme 1: Deleting all amplitudes involving the core virtual}
We make use of three additional schemes to address the numerical 
instabilities discussed previously. The first, which we refer 
as Scheme 1 (S1), is that proposed Zheng et al.\cite{Zheng2019}, 
and employed by Lee and Head-Gordon.\cite{Lee2019} This scheme  
simply excludes any amplitude involving the core virtual. Additionally, 
we chose to exclude singles amplitudes that excite the occupied core electron. 
\begin{align*}
    \text{if(}a &= \bar{h}\;\text{or}\;i = h)\;\;\; a_i^a,\; t_i^a = 0 \\
    \text{if(}a &= \bar{h}\;\text{or}\;b = \bar{h})\;\;\; a_{ij}^{ab},\; t_{ij}^{ab} = 0 \\
    \text{if(}a &= \bar{h}\;\text{or}\;b = \bar{h}\;\text{or}\;c = \bar{h})\;\;\; 
                   t_{ijk}^{abc}(c) = 0
\end{align*}
Under these conditions, the ill-behaved amplitudes are removed by design. 
However, by excluding amplitudes that involve the core virtual, 
we are also excluding part of the correlation between the remaining core electron and 
valence electrons, as will become more clear below. The de-excitation amplitudes in
the Lambda equations, solved to obtain CC properties like $\braket{S^2}$,  
are treated in a completely analogous way. Under these constraints, the Lambda 
equations converged to yield to similar $\braket{S^2}$ values than without them, 
but at a much accelerated pace.

\subsection{Scheme 2: Half-occupied core with zero spin-complement amplitude}
To incorporate some of the correlation missing in S1, Scheme 2 (S2) 
allows for the double substitutions involving the core virtual, $\bar{h}$, that 
also promote the occupied electron in the same core orbital, $h$ - these were found 
to be the leading amplitudes for some of the larger well-behaved S0 
calculations. S2 is pleasing in that, even though core substitutions are involved, 
they are all associated with configurations that retain a core occupancy of 1.
\begin{align*}
    \text{if(}a = \bar{h}\;\text{or}\;i &= h)\;\;\; a_i^a,\; t_i^a = 0 \\
    \text{if(}a = \bar{h}\;\text{or}\;b &= \bar{h})\\
    \text{if(}i &\ne h\;\text{or}\;j \ne h)\;\;\; a_{ij}^{ab},\; t_{ij}^{ab} = 0 \\
    \text{if(}a = \bar{h}\;\text{or}\;b &= \bar{h}\;\text{or}\;c = \bar{h})\;\;\; \\
    \text{if(}i &\ne h\;\text{or}\;j \ne h\;\text{or}\;k \ne h)\;\;\; t_{ijk}^{abc}(c) = 0
\end{align*}
As for S1, the CC de-excitation amplitudes are treated in a completely analogous way. 
We found that, in the case of the mixed singlets, allowing for the double 
substitution that generates the spin complement of the reference, 
$a^\dag_{\bar{h}}a_{\bar{t}} a^\dag_t a_h$  with $t$ being the target orbital,
leads the CC iterations to converge towards the (lower energy) triplet excited 
state, resulting $\braket{S^2}$ values that deviate significantly from 1. Therefore, an 
additional constraint was placed on calculations for the mixed singlet: 
the amplitude associated with said excitation is also set to zero. 
This helped ensure that the $\braket{S^2}$ value of the CCSD wave 
function remained close to 1, signifying that it is a mixed spin configuration. 
Therefore, as with S1, the spin contamination is removed by evaluating the 
singlet energy via Yamaguchi's AP expression.

\subsection{Scheme 3: Half-occupied core with unit spin-complement amplitude}

As a final scheme, and exclusively for the calculations on the mixed 
singlet state, we propose to incorporate all of the conditions 
of S2 but, instead of neglecting the double substitution amplitude, 
$a^\dag_{\bar{h}}a_{\bar{t}} a^\dag_t a_h$, associated 
with the spin complement of the reference, we set it to 1.0; we refer to this as 
Scheme 3 (S3). These conditions force the CC iterations to 
look for the pure singlet starting from the mixed reference. 
As previously, the exact same S3 conditions are imposed on 
the de-excitation amplitudes for the left eigenvectors 
of the similarity transformed Hamiltonian. We found that 
the lambda equations were able to converge even when the 
de-excitation amplitude associated with the spin complement is not forced to be 1.0. 
Enforcing said condition accelerated the convergence to result 
in the same value for $\braket{S^2}$ An attractive 
feature of S3, as will be elaborated on in the Results 
section, is that it bypasses the need for AP altogether because 
the resulting states have $\braket{S^2}$ values relatively close 
to 0. 
S3 is, in fact, similar in spirit to the the bi-configurational 
MR-CC model proposed by Oliphant and Adamowicz in 1991,
\cite{Oliphant1991} (see also the two-determinant Hilbert-space MR-CC,\cite{Kucharski1991, Balkova1992} 
recently employed by Matthews \cite{Matthews2020} in conjunction with 
an orbital-optimized CSF for core excited states). 
However S3 is dramatically simpler because additional 
triple and quadruple excitations that are necessary in MR-CC 
(in order to account for the single and double excitations on top 
of the ``secondary reference'') are omitted here.

The amplitude of the spin complement can also be set to -1.0 
to access the M$_s$ = 0 triplet. This allows us to asses the 
reliability of S3 by comparing its calculated triplet, M$_S$ = 0 
numbers against the M$_s$ = $\pm$1 triplet numbers obtained via S2. 
In the absence of spin-orbit coupling or external magnetic fields, the M$_s$ = 1 and 
M$_s$ = 0 triplet states should be degenerate, so any differences 
reflect the failures of S3 with respect to S2. Naturally, one source of error 
will be the fact that, in S3, the correlation methods treat each individual 
configuration of the CSF unequally.

\section{Results and discussion}

Before discussing the correlated methods, it is worth revisiting the 
spin-pure open shell singlet HF results (labeled as ROKS(HF), as this 
can be viewed as a special case of OO-DFT\cite{hait2021orbital}). 
For the excitations considered, ROKS(HF) achieves a mean absolute 
error (MAE) and RMSE of 0.43 and 0.52 eV. All of the excitations 
involving carbon and nitrogen, and the O 1s - $\sigma^*$/Rydberg 
transitions are overestimated. All of the fluorine and neon 
excitations , and the O 1s - $\pi^*$ transitions are underestimated. 
This element-dependent error distribution with respect to experiment 
leads to a relatively small mean signed error (MSE) of 0.18 eV. 
Using ROKS with the standard SCAN functional,\cite{sun2015strongly} 
the best-performing functional according to a recent study, 
reduces the MAE to 0.16 eV, the RMSE to an impressive 0.19 eV,\cite{Hait2020-core_excitations} 
and an MSE of only -0.08 eV. How well can CC methods limited to 
double or perturbative triple substitutions compete with these results?

We begin the analysis by noting that the FC-CVS-EOM-CCSD aproach 
cannot match ROKS(SCAN), and in fact it scarcely outperforms the 
simple ROKS(HF): FC-CVS-EOM-CCSD achieves an MAE and RMSE of 0.34 
and 0.41 eV. FC-CVS-EOM-CCSD tends to underestimate the excitations 
out of carbon, with an overestimation of 0.34 eV for the 
\ce{\textbf{C}H3OH} 1s $\longrightarrow$ 3s transition being the 
only serious exception. All other excitations are overestimated 
by FC-CVS-EOM-CCSD, except for the \ce{N2} 1s $\longrightarrow$ $\pi^*$ 
and \ce{Be} 1s $\longrightarrow$ 2p excitations, which are underestimated 
by 0.25 and 0.68 eV, respectively. The latter might be a failure of the 
FC-CVS model.

\begin{figure}
  \includegraphics{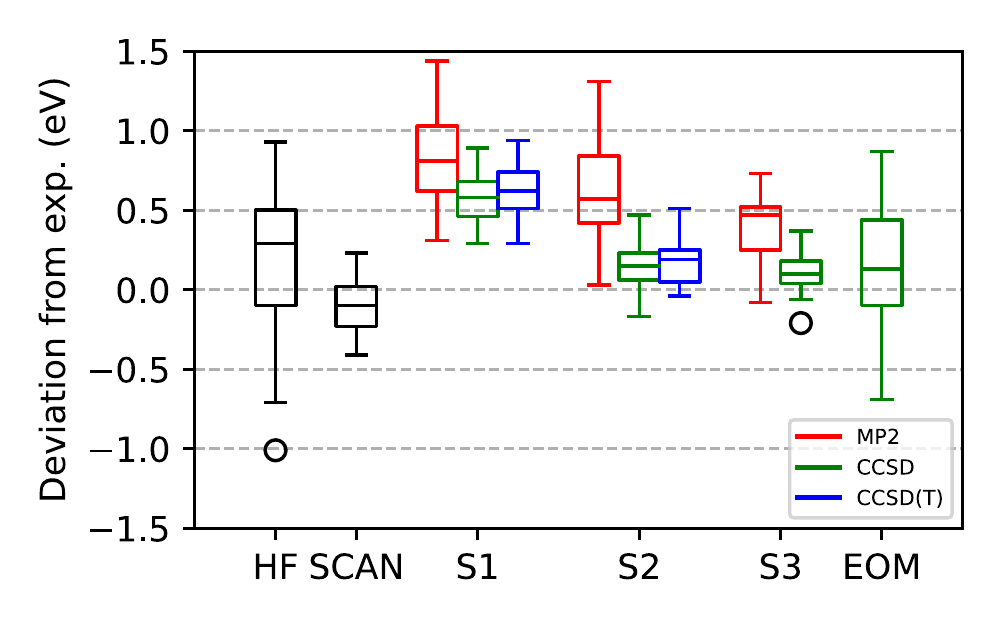}
  \caption{Statistical summary of the accuracy of calculated K-shell core excitations relative to experimental values for the 21 transitions shown in Table \ref{excitations}, as evaluated by ROKS(HF), ROKS(SCAN), the correlated $\Delta$ methods (Schemes S1, S2 and S3), and FC-CVS-EOM-CCSD-EE. For the S1 and S2 approaches, in addition to CCSD itself, the corresponding MP2 and CCSD(T) values are also shown. The specific values corresponding to these statistics are given in Table \ref{excitations} and the Supplementary Information.}
  \label{fgr:excitations}
\end{figure}

In regards to the correlated $\Delta$ methods, addressing the offending 
denominators, either by eliminating all excitations into the core 
virtual (S1) or including only those that retain a core occupancy of 1 
(S2 and S3) resulted in well-behaved, monotonically convergent CC 
calculations in all cases. Furthermore, for Schemes S1 and S2, 
the MP2, CCSD, and CCSD(T) correlation energies of the excited 
states, and the calculated excitation energies seem to converge 
monotonically towards a well defined BSL.

As observed in Figure \ref{fgr:excitations} and the SI, correlated 
calculations via Scheme S1 always overestimate the excitation energy. 
$\Delta$MP2(S1), $\Delta$CCSD(S1), and $\Delta$CCSD(T)(S1) achieve 
MAEs of 0.82, 0.58, 0.63 eV, and RMSEs of 0.88, 0.60, 0.65 eV. 
$\Delta$CCSD(S1) attenuates the most severe failures of 
$\Delta$MP2(S1) - where it overestimates experiment by more than 1 eV:
\ce{H2\textbf{C}O} 1s $\longrightarrow$ $\pi^*$,
\ce{H\textbf{C}N} 1s $\longrightarrow$ $\pi^*$,
\ce{HC\textbf{N}} 1s $\longrightarrow$ $\pi^*$,
\ce{N2} 1s $\longrightarrow$ $\pi^*$, and
\ce{F2} 1s $\longrightarrow$ $\sigma^*$. 
These are all cases where $\Delta$MP2(S1) changes the ROKS(HF) 
results the most - in all cases for worse - with \ce{F2} 
having the largest change in magnitude, at 2.3 eV. $\Delta$CCSD(T)(S1), 
more often than not, seems to very slightly increase the error against experiment 
when compared to $\Delta$CCSD(S1). Including correlation via 
S1, either via MP2, CCSD, or CCSD(T) only decreases the calculated 
values relative to $\Delta$HF in roughly half the cases. The 
MSEs for all the correlated methods under S1 are identical 
to their MAEs, which is is consistent with a systematic 
overestimation of the excitation energies or, conversely, 
an under-correlation of the excited states. Since the results 
are expected to be well near the BSL, and the perturbative 
triples correction changes the CCSD results by a small amount, 
we attribute this to the configurations excluded from the 
correlation treatment for the sake of proper convergence. 

As proposed in the previous section, not all configurations 
involving excitations into the core virtual need to be excluded 
for a safe convergence of the iterative CC procedure. Figure 
\ref{fgr:excitations} and SI show that including some of the 
missing configurations via scheme S2 indeed reduces the error relative 
to S1. $\Delta$MP2(S2), $\Delta$CCSD(S2), and $\Delta$CCSD(T)(S2) 
achieves MAEs of 0.62, 0.18, and 0.20 eV, and RMSEs of 0.69, 0.22 
and 0.25 eV. A small systematic overestimation remains, as suggested 
by MSEs of 0.61, 0.16, and 0.20 eV. Two relevant statistical observations 
are that $\Delta$MP2(S2) still fails to offer an improvement 
over ROKS(HF), and that the (T) correction slightly worsens the 
$\Delta$CCSD results. We note how the well-behaved excitations 
involving the core account for roughly 0.4 eV of the calculated 
excitation energy, as measured by the statistical differences 
between $\Delta$CCSD(S1) and $\Delta$CCSD(S2). This is in agreement with 
the findings of Zheng \emph{et. al}\cite{Zheng2019} and emphazises 
that, if quantitative agreement is desired, a CVS scheme like S1 is inadequate.

Before discussing the performance of S3 in predicting excitation 
energies, we make some other relevant remarks on the scheme. The 
de-excitation amplitudes were usually converged without any modifications 
to yield a CCSD $\braket{S^2}$ close to 0 (or 2, if the triplet state was 
being targeted). Naturally, it often takes many iterations for these 
amplitudes to respond to the large excitation amplitude in T$_2$. 
Imposing the condition analogous to S3 for the de-excitation amplitudes 
accelerated the convergence, never taking more than 35 iterations without 
DIIS for the cases that we studied. As is noted in the SI, a residual 
deviation from an $\braket{S^2}$ value of 0 remained for all calculations. 
The largest of these deviations was for the \ce{C2H2} 1s $\longrightarrow$ 
$\pi^*$ state, with an $\braket{S^2}$ of 0.069, the average being 0.033. 
We suspect that this might be due to the missing excitations described 
in the discussion of S3.

The spin-forbidden excitations into the triplet M$_s$ = 0 manifold were 
calculated with $\Delta$CCSD(S3) by forcing the amplitude of the spin complement 
of the reference to be -1.0; they are listed in SI. We compared these against 
the triplet M$_s$ = 1 excitation energies as calculated by $\Delta$CCSD(S2). 
The largest deviation was of 0.09 eV for the \ce{H2\textbf{C}O} 
1s $\longrightarrow$ $\pi^*$ state, the average being 0.04 eV. 
The M$_s$ = 0 triplet excitations were higher than the M$_s$ = 1 results for all but one case, 
\ce{Be} 1s $\longrightarrow$ 2p,  where the difference is -0.01 eV. This 
is also consistent with the idea that for the M$_s$ = 0 triplets, as 
for the singlets, we are undercorrelating the excited state due to missing 
excitations. An undercorrelation is not present for the M$_s$ = 1 triplet 
because, aside from any spatial symmetry breaking, this is purely a SR situation
that S2 should be able to address. The triplet numbers, as calculated by 
$\Delta$CCSD(S2), match fairly well with the two experimental numbers 
that we found for these spin-forbidden transitions: 114.3 eV for 
\ce{Be} 1s $\longrightarrow$ 2p and 400.12 eV for 
\ce{N2} 1s $\longrightarrow$ $\pi^*$ \cite{Kramida2009, Shaw1982}. 
$\Delta$CCSD(S2) predicts them to be 114.37 eV and 400.24 eV, 
respectively. The average energy difference between the singlet 
and triplet excited states for the set of molecules studied 
here, as calculated by $\Delta$CCSD(S3), is 0.44 eV. Some cases 
worthy of notice are \ce{Be} 1s $\longrightarrow$ 2p, where the 
splitting is 1.16 eV, and \ce{\textbf{C}O} 1s $\longrightarrow$ $\pi^*$, 
with the largest splitting of all: 1.42 eV. Interestingly, the 
splitting for \ce{C\textbf{O}} 1s $\longrightarrow$ $\pi^*$ is 
only 0.34 eV. Another case of relevance are the two Rydberg excitations
\ce{Ne} 1s $\longrightarrow$ 3s and \ce{Ne} 1s $\longrightarrow$ 3p, 
which have the smallest splittings: 0.06 eV and 0.05 eV, respectively.

In Table \ref{excitations}, we present the calculated excitation 
energies of the singlet excited states for the most successful scheme, 
$\Delta$CCSD(S3), against ROKS(HF), ROKS(SCAN), and FC-CVS-EOM-CCSD-EE
\cite{Vidal2019}. All the statistics provided are compared against 
their most recent and / or accurate experimental 
values. The per-molecule results for the remaining schemes are 
listed in the SI. Overall, $\Delta$CCSD(S3) achieves an MAE and 
RMSE of 0.14 and 0.18 eV. The most challenging excitation for 
this method is \ce{H2\textbf{C}O} 1s $\longrightarrow \pi^*$, 
with an overestimation of 0.37 eV from the experimental value of 
287.98 eV by Remmers \emph{et al.}\cite{Remmers1992}. A small 
systematic overestimation remains, as suggested by a MSE of 0.12 eV. 
The only excitation that $\Delta$CCSD(S3) significantly underestimates 
is \ce{\textbf{C}O} 1s $\longrightarrow$ $\pi^*$, which is below Sodhi 
and Brion's result of 534.21 $\pm$ 0.09 eV by 0.21 eV.\cite{Sodhi1984} 

A recent study that is closely-related to our approach is 
the application of a direct two-determinant (TD) CCSD 
protocol\cite{Kucharski1991} to study core excited 
states.\cite{Matthews2020} This procedure follows the 
$\Delta$CC framework through orbital-optimizing a core 
excited configuration, constructing a CSF, and carrying out 
TD-CCSD on top of it. To address the dangerous denominators, 
an equivalent of our Scheme 2 is employed.\cite{Nooijen1995} 
It is shown that TD-CCSD results have a comparable accuracy to 
the $\Delta$CCSD results reported here, with a MAE of 0.10 
eV and RMSE of 0.11 eV against the Coriani implementation 
of CVS-EOM-EE-CCSDT for the three lowest lying core excitations 
of \ce{HCN}, \ce{CO}, \ce{NH3}, and \ce{H2O}. The $\Delta$CC 
approaches presented in our work have the advantage of halving 
the number of amplitudes as compared to the bi-configurational 
TD-CCSD, by virtue of employing pure SR formalism. Furthermore, 
employing the Scheme of choice to accelerate the convergence 
of the Lambda equations enables calculations of excited 
state properties such as gradients and $\braket{S^2}$.

    It is worth emphasizing that, as noted 
    by Vidal \textit{et al.} in their report,\cite{Vidal2019} 
    the good performance of the FC-CVS relative to the earlier 
    CVS scheme\cite{Coriani2015} is due to a cancellation of errors. 
    The ground state CC wave function is under-correlated by imposing 
    the frozen core approximation, bringing down the excitation energy 
    to better match the experimental value. The 
    CVS scheme of Coriani \textit{et al.}\cite{Coriani2015} 
    includes all excitations for the ground state and decouples 
    the core excited states via projection from the valence states 
    in the EOM component of the procedure. As such, it does not 
    benefit from the error cancellation present in the FC-CVS 
    scheme and despite being preferable on formal grounds, it 
    performs worse when comparing to experiment.
    
\begin{table}
  \caption{BSL estimate of the core excitation energies predicted 
  by the best-performing theoretical methods studied in this project 
  compared against their most recent experimental values. The uncertainties 
  of the experimental values (when provided in the reference) are in parentheses.}
  \label{excitations}
  \begin{tabular}{l|ccccc}
  \hline
  Transition & ROKS(HF) & ROKS(SCAN) & $\Delta$CCSD(S3) & EOM-CCSD & Experiment \\
  \hline
\ce{Be} 1s - 2p &115.37 &115.34 &115.53 &114.79 &115.47 \cite{Kramida2009} \\
\ce{C2H4} 1s - $\pi^*$ &285.27 &284.70 &284.77 &284.68 &284.68 (0.1)\cite{Hitchcock1977} \\
\ce{H2\textbf{C}O} 1s - $\pi^*$ &286.42 &285.74 &285.96 &285.62 &285.59\cite{Remmers1992} \\
\ce{C2H2} 1s - $\pi^*$ &286.40 &285.67 &285.84 &285.55 &285.9 (0.1)\cite{Hitchcock1977} \\
\ce{H\textbf{C}N} 1s - $\pi^*$ &286.98 &286.35 &286.51 &286.07 &286.37\cite{Hitchcock1979} \\
\ce{\textbf{C}O} 1s - $\pi^*$ &288.05 &286.99 &287.46 &286.71 &287.40(0.02)\cite{Sodhi1984} \\
\ce{\textbf{C}H3OH} 1s - 3s &288.91 &288.18 &288.34 &288.26 &287.98\cite{Prince2003} \\
\ce{CH4} 1s - 3p($t_2$) &288.38 &287.96 &288.02 &287.9 &288.00 (0.2)\cite{Schirmer1993} \\
\ce{HC\textbf{N}} 1s - $\pi^*$ &400.00 &399.60 &399.80 &399.74 &399.7\cite{Hitchcock1979} \\
\ce{NH3} 1s - 3s &400.97 &400.42 &400.63 &400.82 &400.66 (0.2)\cite{Schirmer1993} \\
\ce{N2} 1s - $\pi^*$ &401.18 &400.80 &401.02 &400.63 &400.88 (0.02)\cite{Sodhi1984} \\
\ce{NH3} 1s - 3p(e) &402.62 &402.18 &402.41 &402.46 &402.33 (0.2)\cite{Schirmer1993} \\
\ce{H2C\textbf{O}} 1s - $\pi^*$ &530.67 &530.83 &530.86 &531.26 &530.82\cite{Remmers1992} \\
\ce{H2O} 1s - 3s &534.15 &533.84 &534.14 &534.44 &534.0 (0.2)\cite{Schirmer1993} \\
\ce{CH3\textbf{O}H} 1s - 3s &534.16 &533.98 &534.24 &534.64 &534.12\cite{Prince2003} \\
\ce{C\textbf{O}} 1s - $\pi^*$ &533.68 &533.97 &534.00 &534.50 &534.21 (0.09)\cite{Sodhi1984} \\
\ce{H2O} 1s - 3p (b$_2$) &536.03 &535.65 &536.08 &536.21 &535.9 (0.2)\cite{Schirmer1993} \\
\ce{F2} 1s - $\sigma^*$ &681.19 &682.43 &682.41 &683.07 &682.2 (0.1)\cite{Hitchcock1981} \\
\ce{HF} 1s - $\sigma^*$ &687.31 &687.44 &687.76 &688.05 &687.4 (0.2)\cite{Hitchcock1981} \\
\ce{Ne} 1s - 3s &864.75 &865.18 &865.37 &865.54 &865.1 (0.1)\cite{Hitchcock1981} \\
\ce{Ne} 1s - 3p &866.58 &866.96 &867.30 &867.40 &867.29\cite{Muller2017} \\
\hline
MSE &0.15 &-0.09 &0.12 &0.11 & \\
MAE &0.43 &0.15 &0.14 &0.34 & \\
RMSE &0.52 &0.19 &0.18 &0.41 & \\
MAX &1.01 &0.41 &0.37 &0.87 & \\
  \end{tabular}
\end{table}

\begin{figure}
  \includegraphics{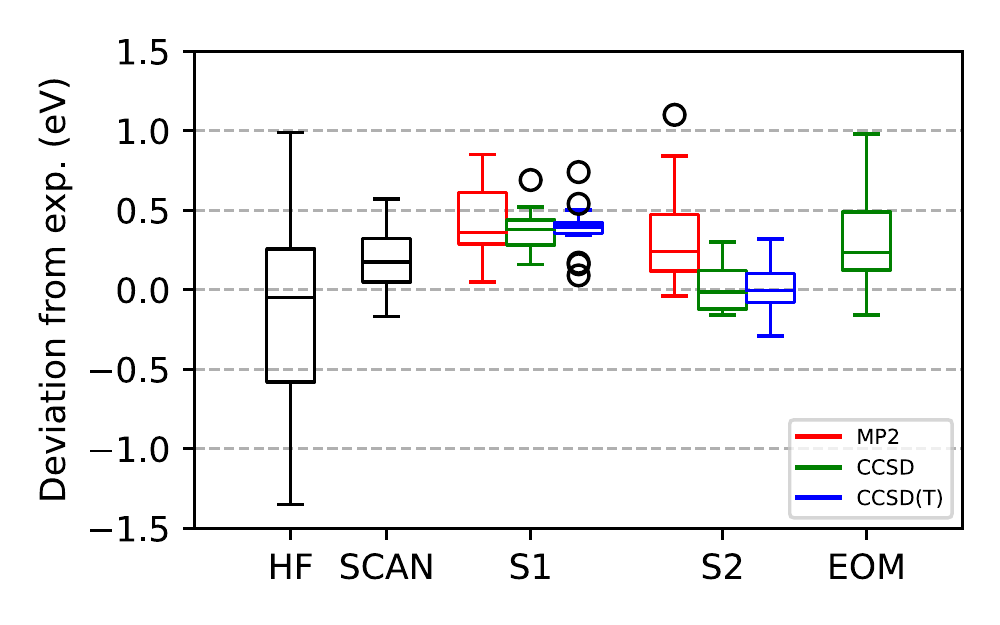}
  \caption{Statistical summary of the accuracy of calculated K-shell core ionizations relative to experimental values for the 18 ionizations shown in Table \ref{tbl:ionizations}, as evaluated by ROHF(SCF), the correlated $\Delta$ methods (Schemes S1 and S2), and FC-CVS-EOM-CCSD-IP. For the S1 and S2 approaches, in addition to CCSD itself, the corresponding MP2 and CCSD(T) values are also shown. The specific values corresponding to these statistics are given in Table \ref{tbl:ionizations} and the Supplementary Information.}
  \label{fgr:ionizations}
\end{figure}

Table \ref{tbl:ionizations} compares the $\Delta$CCSD(S2) core ionizations, 
against those calculated by $\Delta$SCF(HF), $\Delta$SCF(SCAN) and 
FC-CVS-EOM-CCSD-IP. Figure \ref{fgr:ionizations} shows box-whisker plots for 
both the S1 and S2 methods applied to MP2, CCSD, and CCSD(T) relative to 
the same existing methods. The experimental values used as a reference 
are the ones given by Jolly et al.\cite{Jolly1984}, unless a more recent study 
was found. $\Delta$SCF(HF) has a MSE, MAE, and RMSE of -0.15, 0.45, and 0.58 eV, respectively. 
The two most challenging cases for $\Delta$SCF in the ionization data set, \ce{CO} 
and \ce{F2}, are the only cases with an error greater than 1 eV. 
$\Delta$SCF(SCAN) reduces the $\Delta$SCF(HF) errors by more than a factor of two, 
with an MAE and RMSE of 0.21 and 0.25 eV. In contrast to excitations, 
all ionizations except two, \ce{F2} and \ce{Ne}, are overestimated with 
$\Delta$SCF(SCAN), resulting in an MSE similar to its MAE: 0.18 eV. 
The most challenging case for $\Delta$SCF(SCAN) is Be, over 
estimated by 0.51 eV. Somewhat surprisingly $\Delta$SCF(HF) 
predicts the Be experimental ionization perfectly.  

The performance of $\Delta$SCF(HF) against the much more sophisticated 
FC-CVS-EOM-IP-CCSD is once again remarkable, with the MAE and RMSE of 
the latter being 0.35 and 0.45 eV. Elaborating on 
the previous discussion on the specific details of the CVS implementation, 
we note that these FC-CVS-EOM-IP-CCSD errors are roughly five times 
smaller than those reported by Liu et al. for the Coriani-style
CVS-EOM-IP-CCSD.\cite{Liu2019} A similiar situation takes place 
within the context of EOM-EE-CCSD for core excitations, as 
reported by Vidal et al.\cite{Vidal2019}

In contrast to excitations, the correlated $\Delta$ methods using the S1 
model manage to slightly improve upon $\Delta$HF for ionization. 
$\Delta$MP2(S1) increases the HF ionization energy in almost all cases, 
and over 1 eV in several of them:
\ce{H2C\textbf{O}},
\ce{CH3\textbf{O}H},
\ce{C\textbf{O}},
\ce{HF},
\ce{F2}, and
\ce{Ne}.
The only case where $\Delta$MP2(S1) decreases the ionization predicted by $\Delta$HF 
is 
\ce{\textbf{C}O}, 
which is also the second most challenging case for $\Delta$HF, right after \ce{F2}. 
The problematic Be is overestimated by 0.81 eV by $\Delta$MP2(S1). Once again, 
$\Delta$CCSD(S1) alleviates the worst cases in $\Delta$MP2(S1). 
\ce{\textbf{C}O} 
is anomalous in that this is the only case where $\Delta$CCSD(S1) significantly 
worsens the $\Delta$MP2(S1) result, and also the only one where the (T) seems 
to significantly improve the result, correcting the $\Delta$CCSD(S1) 
result by 0.17 eV. Overall, the S1 methods result in MAEs and RMSEs of 
0.42, 0.37, 0.38 eV and 0.49, 0.39, 0.41 eV for MP2, CCSD, and CCSD(T). 
As Lubijic\cite{Ljubic2014} noted in their study, 
$\Delta$MP2(S1) seldom warrants the additional cost over $\Delta$SCF and 
neither extending to CCSD or CCSD(T) seems to improve the results to an extent 
that justifies their cost. As for excitations, a consistent 
overestimation of the core ionization energies, as evidenced by the MSEs 
being equal to the MAEs for all the S1 correlated methods, hints at the 
configurations neglected by the S1 scheme being important. 

Indeed, the improvement in calculated core ionization energies provided 
by the correlated methods under model S2, 
relative to S1, is even more dramatic than it is for the excitations. 
In contrast with $\Delta$MP2(S1), $\Delta$MP2(S2) manages to somewhat 
improve the statistics from $\Delta$HF, bringing down the MAE and RMSE to 
0.33 and 0.44 eV. S2 improves the S1 results for MP2 in almost all cases, 
the only significant exception 
being Be, where $\Delta$MP2(S2) performs the worst: an overestimation of 1.1 eV. 
As with S1, $\Delta$CCSD(S2) alleviates the failures of $\Delta$MP2(S2) 
(significantly for Be) and brings the MAE and RMSE down to 0.12 and 0.15 eV. 
$\Delta$CCSD(T) slightly worsens the statistics by bringing the MAE and RMSE 
to 0.13 and 0.17 eV. The RMSE for $\Delta$CCSD(S2) is more than 2.5 times smaller than for FC-CVS-EOM-IP-CCSD.

The results presented here are comparable to those in Table 5 of Zheng \textit{et al.}
\cite{Zheng2019} The differences can be associated with the different basis 
sets used and the way we are treating the correlation associated with the core virtual.
Whereas in their study, they make estimates to the correlation missing due to freezing 
the core orbital completely (S1) by carrying out unconstrained (S0) calculations 
with denominator thresholds, S2 recovers it by a well-defined protocol. 
 
\begin{table}
  \caption{BSL estimate of the core ionization energies predicted 
  by the best-performing theoretical methods studied in this project 
  compared against their most recent experimental values. The uncertainties 
  of the experimental values (when provided in the reference) are in parentheses.}
  \label{tbl:ionizations}
  \begin{tabular}{l|ccccc}
  \hline
  Transition & $\Delta$SCF(HF) & $\Delta$SCF(SCAN) & $\Delta$CCSD(S2) & EOM-CCSD & Experiment \\
  \hline
\ce{Be} 1s - ion &123.35 &123.92 &123.65 &123.49 &123.35 \cite{Kramida2009} \\
\ce{C2H4} 1s - ion &290.71 &290.92 &290.72 &290.95 &290.88\cite{Jolly1984} \\
\ce{CH4} 1s - ion &290.86 &290.92 &290.69 &290.68 &290.83 \cite{Jolly1984} \\
\ce{C2H2} 1s - ion &291.39 &291.47 &291.21 &291.26 &291.14 \cite{Jolly1984} \\
\ce{\textbf{C}H3OH} 1s - ion &292.63 &292.63 &292.44 &292.52 &292.3 (0.2)\cite{Hempelmann1999} \\
\ce{H\textbf{C}N} 1s - ion &293.76 &293.68 &293.43 &293.34 &293.50\cite{Jolly1984} \\
\ce{H2\textbf{C}O} 1s - ion &294.91 &294.75 &294.50 &294.70 &294.35\cite{Remmers1992} \\
\ce{\textbf{C}O} 1s - ion &297.23 &296.58 &296.47 &296.43 &296.24\cite{Jolly1984} \\
\ce{NH3} 1s - ion &405.48 &405.70 &405.51 &405.77 &405.52\cite{Jolly1984} \\
\ce{HC\textbf{N}} 1s - ion &406.74 &406.96 &406.78 &407.10 &406.8\cite{Jolly1984} \\
\ce{N2} 1s - ion &410.21 &410.15 &409.99 &409.89 &409.9\cite{Jolly1984} \\
\ce{CH3\textbf{O}H} 1s - ion &538.43 &539.08 &538.90 &539.64 &539.06 (0.2)\cite{Hempelmann1999} \\
\ce{H2C\textbf{O}} 1s - ion &538.51 &539.47 &539.29 &540.28 &539.30\cite{Remmers1992} \\
\ce{H2O} 1s - ion &539.49 &539.96 &539.82 &540.29 &539.92\cite{Jolly1984} \\
\ce{C\textbf{O}} 1s - ion &541.79 &542.65 &542.43 &543.10 &542.57 \cite{Jolly1984} \\
\ce{HF} 1s - ion &693.62 &694.30 &694.25 &694.80 &694.0\cite{Jolly1984} \\
\ce{F2} 1s - ion &695.36 &696.54 &696.58 &697.58 &696.71\cite{Jolly1984} \\
\ce{Ne} 1s - ion &869.54 &870.21 &870.31 &870.49 &870.33\cite{Muller2017} \\
\hline
MSE &-0.15 &0.18 &0.02 &0.31 & \\
MAE &0.45 &0.21 &0.13 &0.35 & \\
RMSE &0.58 &0.25 &0.17 &0.45 & \\
MAX &1.35 &0.57 &0.33 &0.98 & \\
  \end{tabular}
\end{table}

\section{Conclusions}

We have studied the use of core-hole orbital-optimized references in SR correlated 
methods to describe core excited and core ionized states of 18 small 
closed-shell organic molecules, and compared them against two of the 
most successful approaches so far: ROKS(SCAN) and FC-EOM-EOM-CC. 
The use of three different schemes (S1, S2, S3) to address 
the convergence problems of the CC equations, and the spin contamination 
of the excited states, were employed. S1 excludes all amplitudes involving 
the half-occupied core orbital associated with the excitation or ionization. 
S2 allows for the ones that retain a core occupancy of 1. 
S3, exclusively for CCSD core excitations, fixes the $T_2$ amplitude 
associated with the spin compliment of a spin symmetry-broken core excited 
reference to $\pm 1.0$, thereby ensuring the proper reference 
CSF is present in the cluster expansion. As evidenced by the 
energetic difference between the singlet and the triplet core 
excited states, addressing the spin contamination associated 
with using a symmetry broken reference is essential for 
quantitative studies using the correlated $\Delta$ methods unless Rydberg 
states are being targeted. 

To compare with experimental core excitations and ionizations 
requires careful attention to basis set convergence, which we 
have addressed by using the aug-cc-pCVXZ basis set for heavy 
atoms (n = T, Q, with extrapolation), and aug-cc-pVDZ for 
hydrogen. With this protocol, $\Delta$CCSD(S3) performs the best 
among the correlated $\Delta$ methods for core excitations, reaching 
an MAE and RMSE of 0.14 and 0.18 eV for CCSD. These statistics 
are on par with the most successful orbital-optimized DFT approach, 
ROKS(SCAN). $\Delta$CCSD(S2) follows closely behind, with an MAE 
and RMSE of 0.18 and 0.22 eV. As such, $\Delta$CCSD with either S2 
or S3 roughly halves the errors of FC-CVS-EOM-CCSD-EE. A similar 
situation takes place for ionizations, where S2 in conjunction with 
CCSD performs the best, by achieving a MAE and RMSE of 0.13 and 
0.17 eV, respectively. $\Delta$CCSD(S2) reduces the 
FC-CVS-EOM-CCSD-IP error by more than a factor of 2.5, and 
outperforms $\Delta$SCF(SCAN), which has an MAE and RMSE of 
0.21 and 0.25 eV. 

The use of a CVS scheme like S1 for the correlated $\Delta$ methods 
is discouraged, if quantitative agreement is sought after. Furthermore, 
as has previously been concluded by others,\cite{Ljubic2014} we cannot 
recommend the use of $\Delta$MP2 for the prediction of core excitations 
or ionizations. In the future, it may be interesting to explore 
whether regularization or further orbital re-optimization
\cite{shee2021regularized, Lee2019} can address the limitations of 
$\Delta$MP2. Finally, we note that the use of the perturbative (T)
triples correction with the best scheme that allows for it, S2, does 
not seem to offer a significant improvement over CCSD. Perhaps this 
is because the effect of triples is small (based on the excellent 
results obtained with $\Delta$CCSD(S2) and $\Delta$CCSD(S3)) or 
perhaps a full triples treatment is needed to obtain further 
significant improvement. 

    There are additional sources for the disagreement with 
    regards to experimental values. Difficulties in measuring 
    XAS spectra often result in slightly different experimental 
    values from different sources (refer to Ref. \citenum{Muller2017} 
    for example) which are often on the order of the errors 
    observed here. We have made our best effort to obtain the 
    most recent and reliable information available at the moment. 
    Additionally, physical effects lacking in our model may also
    contribute to a disagreement with the experiment. There are 
    two such effects that we expect to be of relevance. The first 
    is the fact that we are treating core excited states as 
    formally bound, whereas in reality they are resonances coupling 
    with the Auger continuum.\cite{Carravetta1991} Said effect is
    expected to shift the energy of the resonance. The second is that 
    we are computing vertical excitation energies - a more 
    complete model would incorporate vibronic effects.
    \cite{Coreno1999, Prince1999, DeSimone2002, Duflot2003}

Despite its shortcomings, the main tool for 
routine calculation of XAS is TD-DFT. Furthermore, due to the 
recent advances in LR-DFT-based theory\cite{Hait2022, Park2022} 
the efficient implementations of $\Delta$SCF methods,\cite{Ehlert2020} 
and specialized basis sets\cite{Ambroise2019}, techniques based on 
mean field approaches will likely remain the workhorses for the 
calculation of 
core spectra. Nonetheless, considering an accuracy of less than 0.2 eV 
attained by the $\Delta$CCSD schemes S2 and S3 for specific 
transitions, we expect these to be a promising method for 
providing benchmark theory-based core excitation / ionization 
numbers. Furthermore, the $\Delta$CCSD methods presented here retain the 
formal $\mathcal{O}(N^6)$ scaling of CCSD. Therefore, the hard limit due to 
computational resources on the size of the systems that can be 
tackled by $\Delta$CCSD is equivalent to that of standard SR CCSD. 
The challenges to making $\Delta$CCSD a practical method for the 
calculation of excitation energies, as can now be done with 
EOM-CCSD, is largely implementational. Specialized and efficient 
amplitude windowing algorithms to carry out the particular 
$\Delta$CC scheme and a robust workflow that allows for the $\Delta$CCSD 
calculation on a number of states of interest (which can be carried 
out in parallel)\cite{Ehlert2020} could eventually lead to routine 
$\Delta$CCSD calculations for transition energies. Furthermore, the 
question of compact and efficient basis sets for these orbital-optimized,
wave-function-based correlated calculations deserves future attention. 
New developments for the calculation of transition properties, such 
as oscillator strengths, within the $\Delta$CCSD framework are still 
needed in order to make this approach an attractive alternative to
conventional CC methods for excited states. As a final outlook, 
we point out to the possibility of employing these accurate 
core excited and core-ionized SR CC reference states as the 
starting point for EOM calculations to open up new 
avenues for investigating satellite peaks - formally higher 
excited states beyond the reach of the traditional EOM formalism - 
in core spectroscopies.\cite{Couto2020}

\begin{acknowledgement}

The authors thank Diptarka Hait for fruitful discussions. 
J. L. thanks David Reichman for support. This work was 
supported by the Director, Office of Science, Office 
of Basic Energy Sciences, of the U.S. Department of 
Energy, under Contract No. DE-AC02-05CH11231.

\end{acknowledgement}

\begin{suppinfo}

All the Supporting Information is provided as .xlsx spreadsheets. For both 
excitations and ionizations via the three $\Delta$CC schemes, the following 
data is provided: SCF energies, correlation energies, SCF and CCSD $\braket{S^2}$ values. 
For the ROKS(SCAN) and FC-CVS-EOM-EE-CCSD and FC-CVS-EOM-IP-CCSD calculations, 
the excitation energies are provided.

\end{suppinfo}

\bibliography{achemso-demo}

\end{document}